\documentclass[runningheads]{llncs}
\usepackage{graphicx}
\usepackage[hyphenbreaks]{breakurl}
\usepackage[hyphens]{url}
\usepackage{moreverb,url}

\usepackage[colorlinks,bookmarksopen,bookmarksnumbered,citecolor=black,urlcolor=black]{hyperref}

\newcommand\BibTeX{{\rmfamily B\kern-.05em \textsc{i\kern-.025em b}\kern-.08em
		T\kern-.1667em\lower.7ex\hbox{E}\kern-.125emX}}

\usepackage{graphicx}

\usepackage{smartdiagram}
\smartdiagramset{set color list={
		gray!30,
		gray!30,
		gray!30,
		gray!30,
		gray!30,
		gray!30,
	},
	sequence item border color=black,
	sequence item font size=16,
}

\usepackage{float}

\usepackage{color}
\usepackage{colortbl}

\usepackage{amssymb}
\usepackage{comment}
\usepackage[multiple]{footmisc}
\usepackage{amsmath}

\usepackage{lipsum}
\usepackage{pfnote}
\usepackage{multirow}

\usepackage{pgfplots}
\usepackage{tikz}
\usetikzlibrary{arrows,decorations.pathmorphing,fit,positioning}
\usepackage{xcolor}
\usetikzlibrary{patterns}
\usepackage{amstext}

\usetikzlibrary{mindmap,trees}

\usepackage{fixfoot} 
\DeclareFixedFootnote{\myfnone}{one}  

\usepackage{enumitem}

\usepackage{colortbl}

\usepackage{bbding} 

\definecolor{maroon}{cmyk}{0,0.87,0.68,0.32}

\definecolor{brightgreen}{rgb}{0.4, 1.0, 0.0}

\usetikzlibrary{positioning,calc}

\usepackage{tikz}
\usetikzlibrary{trees}

\tikzset{level 1/.style={level distance=2cm, sibling distance=10cm}}
\tikzset{level 2/.style={level distance=2cm, sibling distance=3cm}}

\tikzset{bag/.style={text width=10em, text centered,yshift=-0.5cm}}

\usepackage{smartdiagram}
\usesmartdiagramlibrary{additions}

\usetikzlibrary{positioning,calc}

\usetikzlibrary{arrows,shapes,positioning,shadows,trees}

\tikzset{
	basic/.style  = {draw, text width=4cm, drop shadow, font=\sffamily, rectangle},
	root/.style   = {basic, rounded corners=2pt, thin, align=center,
		fill=green!30},
	level 2/.style = {basic, rounded corners=6pt, thin,align=center, fill=green!60,
		text width=2em},
	level 3/.style = {basic, thin, align=left, fill=pink!60, text width=2em}
}

\def\addlegendimage{\csname pgfplots@addlegendimage\endcsname}

\begin{document}
\title{Political Popularity Analysis in Social Media}
%
%
\author{Amir Karami\inst{1}\orcidID{0000-0003-1936-7497} \and
	Aida Elkouri\inst{1} }
\authorrunning{A. Karami and A. Elkouri}
%
\institute{University of South Carolina, Columbia SC 29208, USA \\
	\email{karami@sc.edu}, \email{aelkouri@email.sc.edu}\\
	}
\maketitle              
\begin{abstract}
Popularity is a critical success factor for a politician and her/his party to win in elections and implement their plans. Finding the reasons behind the popularity can provide a stable political movement. This research attempts to measure popularity in Twitter using a mixed method. In recent years, Twitter data has provided an excellent opportunity for exploring public opinions by analyzing a large number of tweets. This study has collected and examined 4.5 million tweets related to a US politician, Senator Bernie Sanders. This study investigated eight economic reasons behind the senator's popularity in Twitter.  This research has benefits for politicians, informatics experts, and policymakers to explore public opinion.  The collected data will also be available for further investigation.

\end{abstract}
\section{Introduction}

Social media play an important role in politics and people show their political Internet activity by posting and sharing their opinions \cite{najafabadi2018hacktivism}. This communication technology has been bringing more citizens into the political process and has provided a personal accessible level through the posted political information \cite{kushin2010did}. For example, the percentage of US adults got news from social media has increased from 49\% in the 2012 US election to 62\% in the 2016 US election \cite{pew2016socmedplat}.
Considering the impact of social media on their public's impression, politicians have utilized this new communication technology \cite{hong2012candidates}. 

Twitter with 80 million US users has been considered as one of the top social media platforms. For instance, former president of Chile has asked the members of his cabinet to use Twitter \cite{econtwitter} and Hillary Clinton has officially announced her campaign in Twitter \cite{huffingtonpostclintontwitter}. More than 80 million US Twitter users is a great motivation for local and regional campaigns to analyze tweets \cite{timesnapchat}. Most politicians have a Twitter account and many have a social media team to manage their Twitter account. For example, Barack Obama had a team with 100 staff to work on his social media such as Twitter during his campaign \cite{hong2013benefits}. Besides, there is a new trend that politicians such as Donald Trump have started writing their tweets themselves to have more exciting and informal communications \cite{timetwitterternd2017}.

Public opinion poll is an essential tool in politics. To collect data measuring public opinion, traditional opinion polls use different methods such as face-to-face interview, and phone interview \cite{bbc}. However, the conventional approaches are labor-intensive and time-consuming.
Social media with millions of users and messages per day is a big focus group to mine public opinion \cite{brandwatch1}. Among social media, Twitter with millions of tweets per day has provided a cost-effective data access platform for collecting millions of tweets containing feelings and opinions to facilitate social media research \cite{sewell2013following}. Twitter data has been used in different political applications election analysis \cite{karami2018mining} and non-political applications such as business \cite{he2017examining}, libraries \cite{collins2018social,karami2018us}, social bot analysis \cite{kitzie2018life}, and health like analyzing diabetes, diet, obesity \cite{karami2018characterizing2,shaw2017computational}, exercise \cite{shaw2019anexploratory}, LGBT health \cite{webb2018characterizing,karami2018characterizing}. However, this data has not been considered for popularity analysis.

The popularity of a politician is an critical success factor for the politician and her/his party to win in elections and implement their plans. Finding the reasons behind the popularity can provide a stable political movement. This research investigates Twitter data using computational methods to understand the most important reasons behind a politician's reputation. For our case study, we selected a popular US politician, Bernie Sanders \cite{huffpostpollsanders}. He received the highest amount of small donations from American people in the 2016 US presidential primary election and his campaign has raised more money than Donald Trump's campaign \cite{bbcusfunding}. Although our approach can detect different reasons behind a politician's popularity, we focus on economic issues, as it was the most important issue for the 2016 US voters \cite{pew2016topissues}.

\section{Related Work}

The fast growth of Twitter and its large-scale public available have drawn the attention of researchers for political applications of Twitter data in three directions: (1) social movement analysis, (2) election prediction, and (3)election analysis. Two examples of the first direction are exploring the role of social media in organizing protesters \cite{tufekci2012social,bellin2012reconsidering,bonilla2015ferguson} and studying the behavior of protesters in social media and its effect on social movements \cite{valenzuela2012social,valenzuela2013unpacking}. The second direction has adopted quantitative methods to determine the popularity of candidates \cite{tumasjan2010predicting,boutet2012s,gaurav2013leveraging} and find the most popular candidate and predict the elections \cite{bermingham2011using,sang2012predicting}. The third research category attempts to investigate an election at a macro level such as studying the social media strategy \cite{lamarre2013tweeting} or analyzing economic factors \cite{karami2018mining}. 

Although previous studies have provided valuable insights into political processes, there is a need to find the essential reasons behind a politician's popularity. This paper addresses this gap by applying a mixed method on millions of tweets.

\section{Methodology and Results}
This research proposes a popularity analysis framework with four steps using two text mining techniques including sentiment analysis and topic modeling along with qualitative coding.

\subsection{Data Collection}
We used Twitter4j, a Twitter Java API (Application Programming Interfaces), to collect data using four queries: ``@berniesanders," ``bernie AND sanders", ``sanders", and ``\#sanders". The tweets were collected from January 1, 2016 to July 31, 2016. The collected data will be publicly available in the first author's websites\footnote{\url{https://github.com/amir-karami/Sanders-Tweets-Data}}.

\subsection{Sentiment Analysis}
\begin{figure}[H]	
	
	\begin{tabular}{ |c|p{11cm}| }
		\hline
		\raisebox{-\totalheight}{\includegraphics[width=0.05\textwidth]{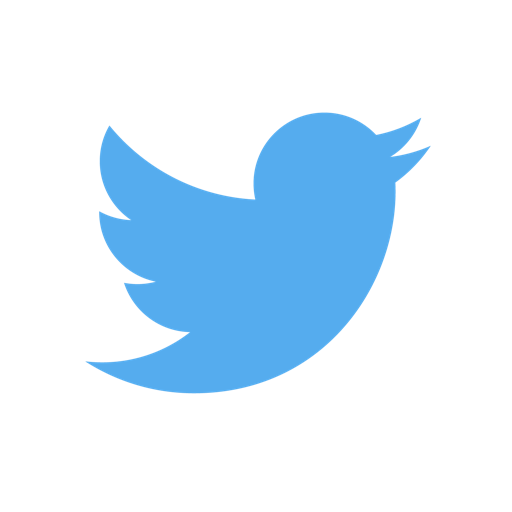}}& \textit{``I agree with Sanders that American can make all public university tuition-free"}\\ \hline	
		\raisebox{-\totalheight}{\includegraphics[width=0.05\textwidth]{twitter.png}}& \textit{``Happy with the candidate who fights for minimum wage"}\\ \hline		 
		
	\end{tabular}
\refstepcounter{table}
	\label{fig:tweet2}
	\caption{A Sample of Positive Tweets}
	
\end{figure}

We used Linguistic Inquiry and Word Count (LIWC) tool \cite{pennebaker2015development} having good sensitivity value, specificity value, and English proficiency measure \cite{golder2011diurnal,karami2014improving1,karami2014exploiting,karami2015online1} for sentiment analysis. Using LIWC, we found 2.1 million positive, 1.7 million negative, and 700,000 neutral tweets. Fig. \ref{fig:tweet2} shows two positive tweets discussing free education and minimum wage. To maintain user privacy, we have lightly edited the represented tweets in this paper to avoid detection.

\subsection{Semantic Analysis}
The third part of our analysis detects main topics discussed by Twitter users during a time frame. Our approach is based on the assumption that people show their support with positive feelings in their tweets. Analyzing a large number of documents like the tweets in our dataset needs computational methods for processing high dimensional data \cite{karami2014fftm,karami2017taming,karami2018computational}. This step applies a topic model to find discussed topics in the detected positive tweets. Latent Dirichlet allocation (LDA) is the most popular and effective general probabilistic topic model to group related words in a corpus \cite{lu2011investigating,karami2015flatm,karami2015afuzzy,karami2018fuzzy}. 

\begin{figure}[ht]
	
	\begin{center}
		\scalebox{0.6}{
			\includegraphics[scale=0.6]{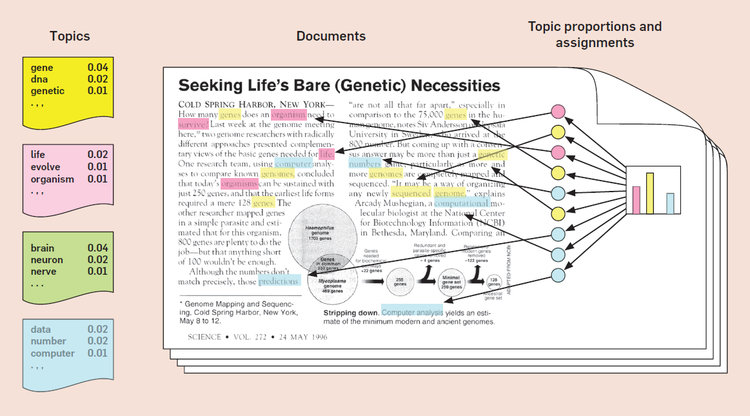}
		}
		\caption{An Example fo LDA \cite{blei2012probabilistic}}
		\label{fig:ldaintuition}
	\end{center}	
\end{figure}

LDA assumes that each tweet in a corpus contains a mixture of topics and each topic is a distribution of the corpus's words \cite{blei2003latent,karami2015fuzzy}. For example, this model assigns ``gene," ``dna," and ``genetic" to a topic with Genetics theme (Figure \ref{fig:ldaintuition}).

After removing the duplicate tweets, retweets, and the tweets containing a URL to retrieve pure personal opinion, we found 307,237 positive tweets. We applied a Java implementation of LDA, Mallet \cite{mallet}, with its default settings and stopwords list to disclose the topics of the 307,237 positive tweets. Using log-likelihood estimation method to identify the optimum number of topics \cite{wallach2009evaluation}, 175 topics were selected as the optimum number of topics.

\subsection{Topic Analysis}

The popularity analysis approach of this research is based on detecting essential reasons. According to the surveys of Gallup and Pew Research Center, the economy was the most critical issue not only in the 2016 election but also in the 2004, 2008, and 2012 US elections \cite{gallup2004topissues,gallup2008topissues,electionsurvey10,pew2016topissues}. In the 2016 US election, Economic was considered in ten dimensions: Jobs \& Income, Trade \& Globalization, Taxes, Entitlement, National Debt, Immigration, Infrastructure, Monetary Policy \& The Federal Reserve, Pay for College, and Minimum Wage. Then we started to the qualitative analysis to identified economic-related topics and labeled them. The authors separately removed nonrelevant topics, either because they were not economic-related topics such as climate change and minorities rights or they were not understandable. By reviewing the top related words such the ones in Table \ref{tab:sample-topics}, we agreed upon assigning single or multiple label(s) based on the ten economic dimensions for each of the relevant topics. For example, we assigned “Minimum Wage” label to a topic containing `Minimum Wage" label to a topic containing \textit{``feelthebern", ``wage", ``support", ``minimum",} and \textit{``workers".}

\begin{table}[ht]
	
	\centering

	\caption{{A Sample of Sanders's Topics}}
	
	\begin{tabular}{ |c|p{3.75cm}|c|c| }
		\hline
		
		\rowcolor{maroon!20} \textbf{Jobs \& Income} & \textbf{Trade \& Globalization}  & \textbf{Taxes}   & \textbf{Entitlement}       \\ \hline
		bernie	& berniesanders   &  berniesanders             & care                    \\
		sanders	& free     &  tax         & universal              \\
		job	    & trade    &  back          & feelthebern       \\
		leverage& increase     &  millions         & people        \\
		economy	& deals      &  taxes            & berniesanders     \\  \hline

		\rowcolor{maroon!20} \textbf{Immigration} & \textbf{Monetary Policy} \&     \textbf{The Federal Reserve} & \textbf{Pay for College}  & \textbf{Minimum Wage }     \\ \hline
		bernie & wall	& free   &  feelthebern                        \\
		sanders & street	& college     &  wage                 \\
		reform & berniesanders 	    & berniesanders    &    supports            \\
		immigration & money& tuition     &  minimum              \\
		good & arguing	& public      &  workers             \\  \hline
		
	\end{tabular}

	\label{tab:sample-topics}

\end{table}

We also explored the distribution of labels to determine the importance of topics for supporters. Table \ref{tab:results} shows that the total weight of the top three reasons, 71\%, was more than the total weight of the rest of the reasons, 29\%. Pay for college, jobs \& income, and entitlement were the top three reasons behind Sanders's popularity. This result confirms the result of a survey plan \cite{sanderedusurvey2016}.

\begin{table}[ht]
	
	\caption{Distribution of Economic Positive Topics}
	\centering
	
	\begin{tabular}{ |c|c|c|} 
		\hline

		\rowcolor{brightgreen} \textbf{Economic Issue}	 &   \textbf{Distribution(\%)} &\textbf{Rank}   \\ \hline 
		\textit{College}                        & 28.8\%&1  \\ \hline
		\textit{Jobs \& Income}	                        & 22.1\% &2 \\ \hline
		\textit{Entitlement}       & 20.3\% &3 \\ \hline
		\textit{Trade \& Globalization}	                   & 8.4\% &4\\ \hline
		\textit{Minimum Wage}                  & 6.8\% &5  \\ \hline
		\textit{Monetary Policy \& The Federal Reserve}        & 6.8\% &5  \\ \hline
		\textit{Taxes}	                                     & 5.1\% &7 \\ \hline
		\textit{Immigration}                              & 1.7\% &8  \\ \hline
		\textit{Infrastructure}                           & 0\% & NA  \\ \hline
		\textit{National Debt}                              & 0\% & NA  \\ \hline

	\end{tabular}
	
	\label{tab:results}
	
\end{table}

The second reason behind the popularity was a plan to raise a national minimum wage . This plan is also in line with traditional surveys \cite{gallupminwage2016,huffingtonpostminwage2017}.  Although jobs \& income and the minimum wage were considered as independent issues, we found an overlap between these two issues. In this case, if we assume that these two reasons represent a single cause, the importance of the combination of these two reasons, 28.9\%, is similar to the importance of the pay for college reason, 28.8\%.
The next reason was entitlement including healthcare and social security that were also in favor of US majority \cite{socialsecunbcwallstjournal2015,gallupmedicare2016}.
Considering the next reason, traditional polls have shown that most Americans were not in favor of the 2016 trade policies and had supported renegotiating major trades \cite{pollingreporttrade2016}. We found that taxes and immigration were the least important reasons for Sander's popularity. This study did not find topics covering national debt and infrastructure issues.

\section{Discussion}

This study applied a mixed method for popularity analysis in social media. There are some key finding informed by this research. First, users don't assign the same weight for all the economic issues. Further, two issues were not among the leading economic concerns of users. Second, college tuition, jobs, and income were the main concerns in the 2016 US election. Third, findings show that the potential of this paper for large-scale social media studies. Fourth, the proposed method can be used with traditional surveys to provide a comprehensive perspective for political events. Fifth, we think that this study has other applications such as analyzing and tracking positive and negative comments for business purposes like the stock market. Finally, the flexibility of the mixed method can help to utilize other computational and qualitative methods.

\section{Conclusion}

This study seeks to the analysis of the economic reasons behind the public's positive feeling. To address the research question, we used a mixed method to develop a popularity analysis approach considering ten economic dimensions. We applied our approach to a massive number of tweets mentioned a popular US politician in 2016 and 2017 to understand the reasons for his popularity. This paper can help politicians, public opinion analysts, knowledge discovery experts, and social scientists to understand people's opinions better.

This study has two limitations. First, the data was collected from one single social media source. Collecting data from other social media such as Facebook can represent more population and opinions. Second, while we considered English tweets, the location of the users was not considered in this analysis. To address these limitations, we will collect data from other social media platforms and consider location of users in our future research.

\section*{Acknowledgements}   
This research is supported in part by the South Carolina Alliance for Minority Participation and the Science Undergraduate Research Fellowships and Exploration Scholars Program programs at the University of South Carolina. All opinions, findings, conclusions and recommendations in this paper are those of the authors and do not necessarily reflect the views of the funding agency.

%
%
%
\bibliographystyle{splncs04}
%
\bibliography{refrence}

\begin{thebibliography}{10}
\providecommand{\url}[1]{\texttt{#1}}
\providecommand{\urlprefix}{URL }
\providecommand{\doi}[1]{https://doi.org/#1}

\bibitem{bellin2012reconsidering}
Bellin, E.: Reconsidering the robustness of authoritarianism in the middle
  east: Lessons from the arab spring. Comparative Politics  \textbf{44}(2),
  127--149 (2012)

\bibitem{bermingham2011using}
Bermingham, A., Smeaton, A.F.: On using twitter to monitor political sentiment
  and predict election results. Sentiment Analysis where AI meets Psychology
  (SAAIP) p.~2 (2011)

\bibitem{blei2012probabilistic}
Blei, D.M.: Probabilistic topic models. Communications of the ACM
  \textbf{55}(4),  77--84 (2012)

\bibitem{blei2003latent}
Blei, D.M., Ng, A.Y., Jordan, M.I.: Latent dirichlet allocation. Journal of
  machine Learning research  \textbf{3},  993--1022 (2003)

\bibitem{bonilla2015ferguson}
Bonilla, Y., Rosa, J.: \# ferguson: Digital protest, hashtag ethnography, and
  the racial politics of social media in the united states. American
  Ethnologist  \textbf{42}(1),  4--17 (2015)

\bibitem{boutet2012s}
Boutet, A., Kim, H., Yoneki, E.: What's in your tweets? i know who you
  supported in the uk 2010 general election. In: The International AAAI
  Conference on Weblogs and Social Media (ICWSM) (2012)

\bibitem{gallup2004topissues}
Carroll, J.: {Economy, Terrorism Top Issues in 2004 Election Vote} (2003),
  \url{http://www.gallup.com/poll/9337/economy-terrorism-top-issues-2004-election-vote.aspx}

\bibitem{collins2018social}
Collins, M., Karami, A.: Social media analysis for organizations: Us
  northeastern public and state libraries case study. Proceedings of the
  Southern Association for Information Systems  (2018)

\bibitem{bbc}
Cowling, D.: {How political polling shapes public opinion} (2015),
  \url{http://www.bbc.com/news/uk-31504146}

\bibitem{gallupminwage2016}
Desilver, D.: 5 facts about the minimum wage. Pew Research Center  (2016),
  \url{http://www.pewresearch.org/fact-tank/2017/01/04/5-facts-about-the-minimum-wage/}

\bibitem{huffingtonpostminwage2017}
Edwards-Levy, A.: Raising the minimum wage is a really, really popular idea.
  The Huffington Post  (2017),
  \url{http://www.huffingtonpost.com/entry/minimum-wage-poll_us_570ead92e4b08a2d32b8e671}

\bibitem{gaurav2013leveraging}
Gaurav, M., Srivastava, A., Kumar, A., Miller, S.: Leveraging candidate
  popularity on twitter to predict election outcome. In: Proceedings of the 7th
  Workshop on Social Network Mining and Analysis. p.~7. ACM (2013)

\bibitem{golder2011diurnal}
Golder, S.A., Macy, M.W.: Diurnal and seasonal mood vary with work, sleep, and
  daylength across diverse cultures. Science  \textbf{333}(6051),  1878--1881
  (2011)

\bibitem{pew2016socmedplat}
Gottfried, J., Shearer, E.: {News Use across Social Media Platforms 2016}
  (2016),
  \url{http://www.journalism.org/2016/05/26/news-use-across-social-media-platforms-2016/}

\bibitem{he2017examining}
He, X., Karami, A., Deng, C.: Examining the effects of online social relations
  on product ratings and adoption: evidence from an online social networking
  and rating site. International Journal of Web Based Communities
  \textbf{13}(3),  344--363 (2017)

\bibitem{hong2013benefits}
Hong, S.: Who benefits from twitter? social media and political competition in
  the us house of representatives. Government Information Quarterly
  \textbf{30}(4),  464--472 (2013)

\bibitem{hong2012candidates}
Hong, S., Nadler, D.: Which candidates do the public discuss online in an
  election campaign?: The use of social media by 2012 presidential candidates
  and its impact on candidate salience. Government Information Quarterly
  \textbf{29}(4),  455--461 (2012)

\bibitem{karami2015fuzzy}
Karami, A.: Fuzzy topic modeling for medical corpora. University of Maryland,
  Baltimore County (2015)

\bibitem{karami2017taming}
Karami, A.: Taming wild high dimensional text data with a fuzzy lash. In: IEEE
  International Conference on Data Mining Workshops (ICDMW). pp. 518--522. IEEE
  (2017)

\bibitem{karami2018mining}
Karami, A., Bennett, L.S., He, X.: Mining public opinion about economic issues:
  Twitter and the us presidential election. International Journal of Strategic
  Decision Sciences (IJSDS)  \textbf{9}(1),  18--28 (2018)

\bibitem{karami2018us}
Karami, A., Collins, M.: What do the us west coast public libraries post on
  twitter? Proceedings of the Association for Information Science and
  Technology  (2018)

\bibitem{karami2018characterizing2}
Karami, A., Dahl, A.A., Turner-McGrievy, G., Kharrazi, H., Shaw, G.:
  Characterizing diabetes, diet, exercise, and obesity comments on twitter.
  International Journal of Information Management  \textbf{38}(1), ~1--6 (2018)

\bibitem{karami2014fftm}
Karami, A., Gangopadhyay, A.: Fftm: A fuzzy feature transformation method for
  medical documents. Proceedings of BioNLP 2014 pp. 128--133 (2014)

\bibitem{karami2015flatm}
Karami, A., Gangopadhyay, A., Zhou, B., Karrazi, H.: Flatm: A fuzzy logic
  approach topic model for medical documents. In: Fuzzy Information Processing
  Society (NAFIPS) held jointly with 2015 5th World Conference on Soft
  Computing (WConSC), 2015 Annual Conference of the North American. pp.~1--6.
  IEEE (2015)

\bibitem{karami2015afuzzy}
Karami, A., Gangopadhyay, A., Zhou, B., Kharrazi, H.: A fuzzy approach model
  for uncovering hidden latent semantic structure in medical text collections.
  iConference 2015 Proceedings  (2015)

\bibitem{karami2018fuzzy}
Karami, A., Gangopadhyay, A., Zhou, B., Kharrazi, H.: Fuzzy approach topic
  discovery in health and medical corpora. International Journal of Fuzzy
  Systems  \textbf{20}(4),  1334--1345 (2018)

\bibitem{karami2018computational}
Karami, A., Pendergraft, N.M.: Computational analysis of insurance complaints:
  Geico case study. International Conference on Social Computing,
  Behavioral-Cultural Modeling, \& Prediction and Behavior Representation in
  Modeling and Simulation  (2018)

\bibitem{karami2018characterizing}
Karami, A., Webb, F., Kitzie, V.L.: Characterizing transgender health issues in
  twitter. Proceedings of the Association for Information Science and
  Technology  (2018)

\bibitem{karami2015online1}
Karami, A., Zhou, B.: Online review spam detection by new linguistic features.
  iConference 2015 Proceedings  (2015)

\bibitem{karami2014exploiting}
Karami, A., Zhou, L.: Exploiting latent content based features for the
  detection of static sms spams. Proceedings of the American Society for
  Information Science and Technology  \textbf{51}(1), ~1--4 (2014)

\bibitem{karami2014improving1}
Karami, A., Zhou, L.: Improving static sms spam detection by using new
  content-based features. The 20th americas conference on information systems
  (AMCIS)  (2014)

\bibitem{kitzie2018life}
Kitzie, V.L., Mohammadi, E., Karami, A.: ``life never matters in the democrats
  mind": Examining strategies of retweeted social bots during a mass shooting
  event. Proceedings of the Association for Information Science and Technology
  (2018)

\bibitem{kushin2010did}
Kushin, M.J., Yamamoto, M.: Did social media really matter? college students'
  use of online media and political decision making in the 2008 election. Mass
  Communication and Society  \textbf{13}(5),  608--630 (2010)

\bibitem{lamarre2013tweeting}
LaMarre, H.L., Suzuki-Lambrecht, Y.: Tweeting democracy? examining twitter as
  an online public relations strategy for congressional campaignsÕ. Public
  Relations Review  \textbf{39}(4),  360--368 (2013)

\bibitem{lu2011investigating}
Lu, Y., Mei, Q., Zhai, C.: Investigating task performance of probabilistic
  topic models: an empirical study of plsa and lda. Information Retrieval
  \textbf{14}(2),  178--203 (2011)

\bibitem{mallet}
McCallum, A.K.: {MALLET: A Machine Learning for Language Toolkit.} (2002),
  \url{http://mallet.cs.umass.edu/topics.php}

\bibitem{najafabadi2018hacktivism}
Najafabadi, M.M., Domanski, R.J.: Hacktivism and distributed hashtag spoiling
  on twitter: Tales of the\# irantalks. First Monday  \textbf{23}(4) (2018)

\bibitem{gallupmedicare2016}
Newport, F.: Majority in u.s. support idea of fed-funded healthcare system.
  Gallup  (2016),
  \url{http://www.gallup.com/poll/191504/majority-support-idea-fed-funded-healthcare-system.aspx}

\bibitem{pennebaker2015development}
Pennebaker, J.W., Boyd, R.L., Jordan, K., Blackburn, K.: The development and
  psychometric properties of liwc2015. UT Faculty/Researcher Works  (2015)

\bibitem{pew2016topissues}
{Pew Research Center}: {Top voting issues in 2016 election} (2016),
  \url{http://www.people-press.org/2016/07/07/4-top-voting-issues-in-2016-election/}

\bibitem{pollingreporttrade2016}
{Polling Report}: International trade / global economy. NBC News/Wall Street
  Journal  (2016), \url{http://www.pollingreport.com/trade.htm}

\bibitem{sanderedusurvey2016}
Pounds, S.: Is college worth it? americans see it as a good investment,
  bankrate survey finds. Bankrate  (2016),
  \url{http://www.bankrate.com/finance/consumer-index/money-pulse-0816.aspx}

\bibitem{timetwitterternd2017}
Rumer, A.: President trump's twitter habit is leading other politicians to pick
  up their smartphones. Time  (2017),
  \url{http://time.com/4822054/donald-trump-twitter-social-media-politicians/}

\bibitem{gallup2008topissues}
Saad, L.: {Iraq and the Economy Are Top Issues to Voters} (2008),
  \url{http://www.gallup.com/poll/104320/iraq-economy-top-issues-voters.aspx}

\bibitem{electionsurvey10}
Saad, L.: {Economy Is Dominant Issue for Americans as Election Nears} (2012),
  \url{http://www.gallup.com/poll/158267/economy-dominant-issue-americans-election-nears.aspx}

\bibitem{socialsecunbcwallstjournal2015}
{Sander Website}: Broad public support for bernieÕs plan to expand social
  security. NBC News/Wall Street Journal  (2015),
  \url{https://berniesanders.com/broad-public-support-for-bernies-plan-to-expand-social-security/}

\bibitem{sang2012predicting}
Sang, E.T.K., Bos, J.: Predicting the 2011 dutch senate election results with
  twitter. In: Proceedings of the Workshop on Semantic Analysis in Social
  Media. pp. 53--60. Association for Computational Linguistics (2012)

\bibitem{sewell2013following}
Sewell, R.R.: Who is following us? data mining a library's twitter followers.
  Library Hi Tech  \textbf{31}(1),  160--170 (2013)

\bibitem{shaw2017computational}
Shaw~Jr, G., Karami, A.: Computational content analysis of negative tweets for
  obesity, diet, diabetes, and exercise. Proceedings of the Association for
  Information Science and Technology  \textbf{54}(1),  357--365 (2017)

\bibitem{shaw2019anexploratory}
Shaw~Jr, G., Karami, A.: An exploratory study of (\#)exercise in the
  twittersphere. iConference 2019 Proceedings  (2019)

\bibitem{brandwatch1}
Smit, K.: {Marketing: 96 Amazing Social Media Statistics and Facts} (2016),
  \url{https://www.brandwatch.com/2016/03/96-amazing-social-media-statistics-and-facts-for-2016}

\bibitem{timesnapchat}
Stein, J.: Why snapchat's snappy. Time  (2017)

\bibitem{econtwitter}
{The Economist}: {Politics and Twitter: Sweet to tweet} (2010),
  \url{http://www.economist.com/node/16056612}

\bibitem{huffpostpollsanders}
{The Huffington Post Pollster}: {Bernie Sanders Favorable Rating} (2017),
  \url{http://elections.huffingtonpost.com/pollster/bernie-sanders-favorable-rating}

\bibitem{bbcusfunding}
Thomas, Z.: {US election 2016: Who's funding Trump, Sanders and the rest?}
  (2016), \url{http://www.bbc.com/news/election-us-2016-35713168}

\bibitem{tufekci2012social}
Tufekci, Z., Wilson, C.: Social media and the decision to participate in
  political protest: Observations from tahrir square. Journal of Communication
  \textbf{62}(2),  363--379 (2012)

\bibitem{tumasjan2010predicting}
Tumasjan, A., Sprenger, T.O., Sandner, P.G., Welpe, I.M.: Predicting elections
  with twitter: What 140 characters reveal about political sentiment. In: ICWSM
  10 (2010)

\bibitem{valenzuela2013unpacking}
Valenzuela, S.: Unpacking the use of social media for protest behavior: The
  roles of information, opinion expression, and activism. American Behavioral
  Scientist  \textbf{57}(7),  920--942 (2013)

\bibitem{valenzuela2012social}
Valenzuela, S., Arriagada, A., Scherman, A.: The social media basis of youth
  protest behavior: The case of chile. Journal of Communication
  \textbf{62}(2),  299--314 (2012)

\bibitem{huffingtonpostclintontwitter}
Velencia, J.: {Hillary Clinton's 2016 Announcement Caused Twitter To Freak Out}
  (2015),
  \url{http://www.huffingtonpost.com/2015/04/13/hillary-clinton-announcement-on-social-media_n_7057020.html}

\bibitem{wallach2009evaluation}
Wallach, H.M., Murray, I., Salakhutdinov, R., Mimno, D.: Evaluation methods for
  topic models. In: Proceedings of the 26th Annual International Conference on
  Machine Learning. pp. 1105--1112. ACM (2009)

\bibitem{webb2018characterizing}
Webb, F., Karami, A., Kitzie, V.: Characterizing diseases and disorders in gay
  users' tweets. Proceedings of the Southern Association for Information
  Systems  (2018)

\end{thebibliography}
\end{document}